# Enhancing Unsupervised Keyword Extraction in Academic Papers through Integrating Highlights with Abstract

Yi Xiang, Chengzhi Zhang[*]

Department of Information Management, Nanjing University of Science and Technology, Nanjing 210094, China

**Abstract:** Automatic keyword extraction from academic papers is a key area of interest in natural language processing and information retrieval. Although previous research has mainly focused on utilizing abstract and references for keyword extraction, this paper focuses on the highlights section—a summary describing the key findings and contributions, offering readers a quick overview of the research. Our observations indicate that highlights contain valuable keyword information that can effectively complement the abstract. To investigate the impact of incorporating highlights into unsupervised keyword extraction, we evaluate three input scenarios: using only the abstract, the highlights, and a combination of both. Experiments conducted with four unsupervised models on Computer Science (CS), Library and Information Science (LIS) datasets reveal that integrating the abstract with highlights significantly improves extraction performance. Furthermore, we examine the differences in keyword coverage and content between abstract and highlights, exploring how these variations influence extraction outcomes. The data and code are available at https://github.com/xiangyi-njust/Highlight-KPE.

## 1. Introduction

Keywords reflect the research problem, methods, and conclusions of a paper, helping readers quickly assess its relevance to their work, thereby reducing time spent on irrelevant research and increasing efficiency. However, not all papers provide keywords (Kim et al., 2010). To address this issue, scholars have proposed automatic keyword extraction, which has been widely applied in scientometric studies, including the construction of thematic networks and knowledge graphs, the identification of research fronts and the analysis of emerging trends (Hou et al., 2024; Wang et al., 2024). Existing approaches to keyword extraction can generally be divided into two categories depending on whether training data are required: supervised (Ercan & Cicekli, 2007; Yan et al., 2024) and unsupervised (Zhang et al., 2023a; Tohalino et al., 2024) methods. Although supervised models generally outperform their unsupervised counterparts, their effectiveness is often constrained by the size and domain specificity of the training data, limiting their generalizability. In contrast, unsupervised methods are more commonly used in academic keyword extraction due to their broader applicability.

The common approach to unsupervised keyword extraction begins by generating a set of candidate keywords from the input text, followed by evaluating their importance based on surface statistical or semantic features. The richness of the input text sets the upper limit for extraction performance—more informative content provides a broader pool of candidates, thereby increasing the chances of capturing relevant keywords. Existing research often utilizes abstract or full texts, with abstract being favored for their concise representation of key research content and easier accessibility compared to full texts (Hulth, 2003). Some researchers have also explored utilizing references or incorporating document structure information for keyword extraction (Zhang et al., 2022a; Zhang et al., 2025). However, the majority of studies emphasize optimizing the evaluation of candidate keywords' importance, with relatively little focus on enriching the input text to enhance performance.

---

[*] Corresponding author, Email: zhangcz@njust.edu.cn

In this study, we focus on a specific type of text found in academic papers: Highlights and explore their integration with abstract for unsupervised keyword extraction. Highlights, typically included in the online version of papers, consist of 3 to 5 sentences provided by the authors, summarizing the paper's key innovations and contributions. Their primary purpose is to help readers quickly understand the research content before downloading the paper and to increase the paper's visibility in search engines, thereby accelerating scientific dissemination[1]. Figure 1 shows the highlights and keyword information from the online version of a paper, with portions of the highlights corresponding to the keywords highlighted in red. As illustrated, highlights are not only concise and focused key points, but they also contain rich keyword information.

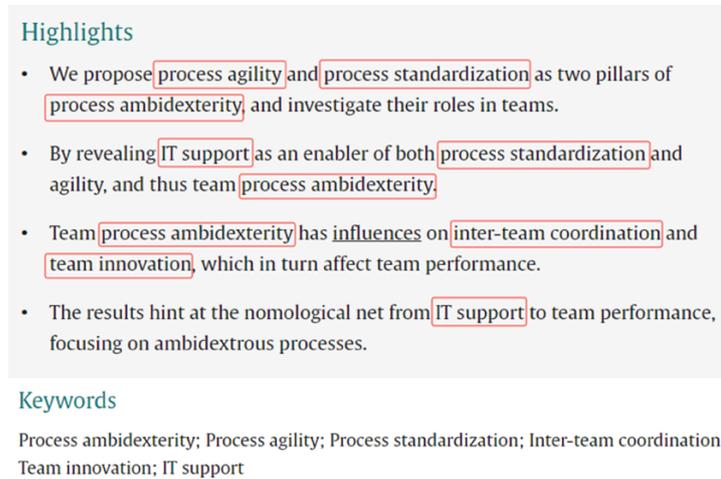

**Fig 1. Example of academic paper highlights. The content enclosed in the red rectangle represents the keyword information within the highlights of the paper[2].**

Given the significant content and functional similarities between highlights and keywords, we propose integrating highlights into academic keyword extraction research. To explore this, we conduct experiments using data from 2,590 papers in LIS field and 2,996 papers in CS field, analyzing abstract, highlights, and keywords. We employ four unsupervised extraction models to assess performance changes with different text inputs. Two methods are tested for integrating information: (1) directly concatenating abstract and highlights, and (2) filtering noise from abstract by evaluating the semantic similarity between their sentences and the highlights, then using the filtered abstract as input. Using these approaches, we generate seven types of text inputs and evaluated their performance across the four models. The results show that simple text concatenation yielded the best performance. To gain deeper insights into the impact of information integration, we examine the differences between abstract and highlights in terms of keyword coverage and content.

The main contributions of this study are summarized as follows:

First, we are the first to explore the integration of highlights with abstract for unsupervised

---

[1] https://www.elsevier.com/researcher/author/tools-and-resources/highlights
[2] https://www.sciencedirect.com/science/article/pii/S0268401219308643

keyword extraction, advancing research on enhancing extraction performance by incorporating external information. This approach also provides valuable insights into the use of highlights in downstream tasks and bibliometric analyses, contributing to a deeper understanding of this knowledge module within academia.

Second, we compare different methods of integrating abstract and highlights information, finding that simple text concatenation significantly improves extraction performance in current unsupervised models, with the order of concatenation having minimal impact on the results.

Third, we validate the effectiveness of filtering abstract sentences based on their semantic similarity to highlights, demonstrating that not all information in an abstract is equally valuable for keyword extraction. Future research should focus on techniques that retain only the most relevant information from abstract.

Finally, we compare abstract and highlights in terms of keyword coverage and content, analyzing the reasons behind performance improvements and variations across datasets using the same input method. This analysis enhances the interpretability of our experimental methods and conclusions, addressing an area that has been relatively underexplored in previous studies.

All data and source code from this study are freely available at the GitHub website: https://github.com/xiangyi-njust/Highlight-KPE.

## 2. Related Work

This study focuses on unsupervised keyword extraction from academic papers, emphasizing the enhancement of extraction performance by integrating highlights information into the abstract. The related research areas include unsupervised keyword extraction methods and the analysis and automatic prediction of highlights content in academic papers.

### 2.1 Unsupervised Keyword Extraction

The research paradigm for unsupervised keyword extraction involves identifying a set of candidate keywords from a text, calculating their importance according to specific rules, and selecting the top k phrases with the highest scores as the extraction results (Song et al., 2023a). The primary focus of such research is on determining how to measure the importance of candidate keyphrases. Existing work has evolved through three stages: traditional statistical-based methods, graph-based methods, and embedding-based methods (Papagiannopoulou & Tsoumakas, 2020). Table 1 lists representative works across these three types of unsupervised keyword extraction methods.

Statistical methods (Jones, 2004; Campos et al., 2018) assess keyphrase importance based on surface features such as frequency and position within the document. Graph-based methods (Bougouin et al., 2013; Florescu & Caragea, 2017; Mihalcea & Tarau, 2004) represent words in the document as nodes in a graph, with edges based on relationships like co-occurrence, and then iteratively compute the importance of each node using graph ranking algorithms until convergence. Although traditional approaches have explored various features and methods for unsupervised keyword extraction, they have limitations in capturing and utilizing semantic information from the text.

Benefiting from the rapid development of text representation learning, embedding-based methods have achieved better performance in unsupervised keyword extraction. These methods leverage language models to obtain high-dimensional semantic representations of phrases and documents. Bennani-Smires et al. (2018) calculated phrase importance scores by directly computing the cosine similarity between phrase and document vectors. Song et al. (2023b) used sentence embedding models and ELMo (Peters et al., 2018) to obtain embeddings better suited for capturing semantic relatedness via cosine similarity, and they introduced a position penalty term to improve extraction performance on long documents. Zhang et al. (2022b) addressed the bias caused by text length mismatches between phrases and documents by masking candidate keyphrases in the document and then inputting both masked and unmasked documents into BERT (Devlin et al., 2019) to compute the cosine similarity between the vectors.

In addition to direct semantic similarity calculations, other approaches have explored deep semantic structures for unsupervised keyword extraction. Tsvetkov & Kipnis (2023) introduced Shannon's information entropy into keyword extraction, proposing an extraction method based on the minimum conditional entropy under a pre-trained language model. Zhang et al. (2023b) considered sentence importance, arguing that keyphrases appearing in important sentences are more likely to be keywords. They developed a hierarchical graph representation model that merges sentence and word importance to score candidate keyphrases. Song et al. (2023b) observed latent hierarchical structures among candidate keywords in syntactic and semantic layers, which are difficult to capture in Euclidean space, and thus proposed HyperRank, a model that jointly captures global and local contextual information in hyperbolic space to estimate keyphrase importance.

Table 1. Related work on unsupervised keyword extraction

| Category | Author | Corpus | Method | Main findings |
|---|---|---|---|---|
| Statistic-based | Jones (2004) | Cranfield, Inspec, Kenn | – | This work differentiates keywords by collection frequency, offering a viable approach widely adopted in subsequent retrieval research. |
| | Campos et al. (2018) | - | Yake | This method ranks candidate terms using aggregate scores from statistical features like frequency, position, and capitalization, then outputs the final keywords. |
| Graph-based | Mihalcea & Tarau (2004) | Inspec | TextRank | This method maps candidate keywords to graph nodes, links nodes by co-occurrence, and iteratively ranks them to identify final keywords. |
| | Bougouin et al. (2013) | Inspec, SemEval, WikiNews. DEFT | TopicRank | This method clusters candidate keywords into topics, uses graph algorithms to select important topics, and then selects keywords from topics. |
| | Florescu & Caragea (2017) | KDD, WWW, Nguyen | PositionRank | This method integrates the position information of candidate words into the PageRank algorithm. |
| Embedding-based | Bennani-Smires et al. (2018) | Inspec, DUC, NUS | EmbedRank | This method selects important words by comparing the semantic similarity between candidate words and documents. |
| | Zhang et al. (2022b) | Inspec, SemEval2010, NUS, DUC2001, SemEval2017, Krapivin | MDERank | This method selects the words with the greatest semantic impact as keywords by comparing the semantic similarity between the document after removing the candidate and original words. |
| | Tsvetkov & Kipnis (2023) | Inspec, SemEval2010, SemEval2017, Inspec | EntropyRank | This method extracts keywords based on the principle of minimizing conditional entropy in pre-trained language models. |
| | Song, et al. (2023b) | DUC2001, SemEval2010 | HyperRank | This method evaluates the importance of candidate keywords by modeling global and contextual information in a hyperbolic space. |

Comparing existing approaches, statistical methods are the simplest to implement but often perform poorly in practical applications, as they rely solely on basic statistical features to determine phrase importance. Graph-based methods are more widely used, particularly in scientometrics and related fields (Bu et al., 2021; Hwang & Shin, 2019; Jiang et al., 2022), and are often employed as baselines for performance comparison with embedding-based methods. Therefore, in this study, we selected two widely used graph-based methods for keyword extraction: TextRank (Mihalcea & Tarau, 2004) and PositionRank (Florescu & Caragea, 2017). TextRank primarily considers term frequency and co-occurrence information, whereas PositionRank further incorporates the positional information of terms within the text. Embedding-based methods, which estimate phrase importance based on semantic relationships within the document, have been shown to substantially improve extraction performance compared to traditional approaches. Accordingly, we included two pre-trained language model-based methods in our experiments—MDERank (Zhang et al., 2022b) and PromptRank (Kong et al., 2023)—and compared their performance with that of the graph-based unsupervised keyword extraction models. Specifically, MDERank measures term importance based on semantic similarity, while PromptRank generates keyword sequences directly through text generation.

## 2.2 Content Analysis and Automatic Prediction of Highlights

Highlights are concise summaries that distill the core contributions of an article, initially popularized in news media to reduce reading time by presenting 3-4 key points related to the main topic (Svore et al., 2007). Highlights serve a similar purpose in academic papers, offering readers a quick overview of the paper's core content. However, because highlights are not typically included in the formally published content of a paper, research specifically focusing on academic highlights is relatively limited. We summarize existing work on highlights in Table 2, which mainly includes studies on content analysis and the automatic prediction of highlights in academic papers.

Yang (2016) analyzed highlights from 240 articles across both soft and hard sciences using text and keyword analysis. The study revealed that authors from different disciplines emphasize different aspects in their highlights: authors in hard sciences tend to focus on methods, while those in soft sciences highlight discussions on phenomena and results. Differences in pronoun usage and the active or passive voice in highlights also reflect the varying approaches to showcasing research outcomes and self-promotion across disciplines. Wang et al. (2023) developed a classification framework for the content described in highlights, categorizing sentences into research introduction, objectives, and background, processes, and methods, conclusions or discussions, and contributions and advocacy. By comparing the distribution of these categories in 239 papers from the fields of library and information science and computer science, they found that the former field emphasizes results, while the latter prioritizes methods. Additionally, library and information science highlight the application of research, whereas computer science highlights its value through comparisons with previous studies.

Since many journals do not provide highlights, some researchers have begun developing automated methods to extract highlights from papers. Collins et al. (2017) constructed the CSPubSum dataset for training models to automatically extract highlights. Wang et al. (2018) applied 23 unsupervised extractive summarization methods to obtain highlights, demonstrating the efficacy of automatic summarization techniques for this task. Cagliero & La Quatra (2020) trained supervised learning models using features such as sentence position and similarity to automatically extract highlights from papers. La Quatra & Cagliero (2022) trained BERT models on datasets like CSPubSum to automatically extract highlights and compared the performance differences when extracting from various sections of a paper, such as abstracts and conclusions. Rehman et al. (2023) employed a generative approach, using a pointer-generator network with coverage mechanisms to generate highlights sentences based on the original paper content. This approach can produce sentences not present in the original text, and comparative results show its advantage over extractive methods.

Table 2. Related work on highlights in academic papers

| Category | Author | Field of Corpus | Size of Corpus | Main Works |
|---|---|---|---|---|
| Content Analysis | Yang (2016) | Arts and Humanities<br>Language and Linguistics<br>Engineering and Technology<br>Medicine | 240 papers | This research compares highlights content emphasis in soft versus hard science papers and assesses author and editor opinions on the utility of highlights through a survey. |
| | Wang et al. (2023) | Computer Science<br>Library and Information Science | 630 papers;<br>239 papers; | This study develops a 5-category framework to analyze the characteristics of highlights sentences finely and compares their distribution in LIS and CS to reveal the differing emphases in scholars' descriptions of paper contributions. |
| Extract Highlight | Wang et al. (2018) | Information Science | 511 papers | This study identified key features for highlights detection and applied a scoring mechanism to extract highlights sentences from the original text. |
| | Cagliero & La Quatra (2020) | Computer Science<br>Artificial Intelligence<br>Bio-medical | 10,281 papers;<br>264 papers;<br>10,760 papers; | This study extracts features from sentences in the paper and applies a traditional machine learning model to identify and output the most salient sentences. |
| | La Quatra & Cagliero (2022) | Computer Science<br>Artificial Intelligence<br>Bio-medical | 10,281 papers;<br>264 papers;<br>10,760 papers; | This method trains a fully connected neural network on BERT embeddings to select sentences closely aligned with the gold standard highlights automatically. |
| Generate Highlight | Rehman et al. (2022) | Computer Science | 10,147 papers | This study incorporates named entity recognition into a pointer-generator network equipped with a coverage mechanism for the automatic generation of academic paper highlights |
| | Rehman et al. (2023) | Computer Science;<br>Mix (Computer Science, Biological, Chemistry, Energy and so on) | 10,142 papers;<br>19,785 papers; | This study uses a pre-trained model to convert sentences into embeddings, processed by a pointer-generator network to generate highlights from the paper autonomously. |

While traditional keyword extraction from academic papers often relies on abstract, this study posits that integrating highlights with abstract can enhance extraction performance. Given that highlights succinctly summarize a paper's methods and conclusions, they likely contain significant keyword information. Additionally, because highlights differ from abstract in content (La Quatra & Cagliero, 2022), they may provide keywords that are absent from abstract. Thus, this study conducts several comparative experiments to assess whether incorporating highlights with abstract improves keyword extraction over using abstract alone. It is important to note that this research only considers papers that include author-provided highlights. Although automatic extraction or generation of highlights from papers has been extensively studied, these methods generally use abstract as input, resulting in generated highlights that closely resemble the abstract.

This similarity could confound our experimental analysis, so only papers with author-provided highlights are considered in this study.

## 3. Methodology

This study focuses on the task of unsupervised keyword extraction from academic papers, specifically exploring whether incorporating highlights into traditional abstract-based extraction methods can improve model performance. The research framework is outlined in Figure 2. Initially, we gather abstract, keywords, and highlights information from the Scopus database and the Elsevier website to construct our experimental dataset. We then evaluate model performance using either the abstract or the highlights as input independently. Building on this, we combined the abstract and highlights texts in various ways to create new inputs and compared the results against those obtained from single-source information. For the keyword extraction models, we select two graph-based methods and two pre-trained language model-based methods. The evaluation and analysis of the results are discussed in Sections 4 and 5. The remainder of this section provides a detailed introduction to the specific keyword extraction process.

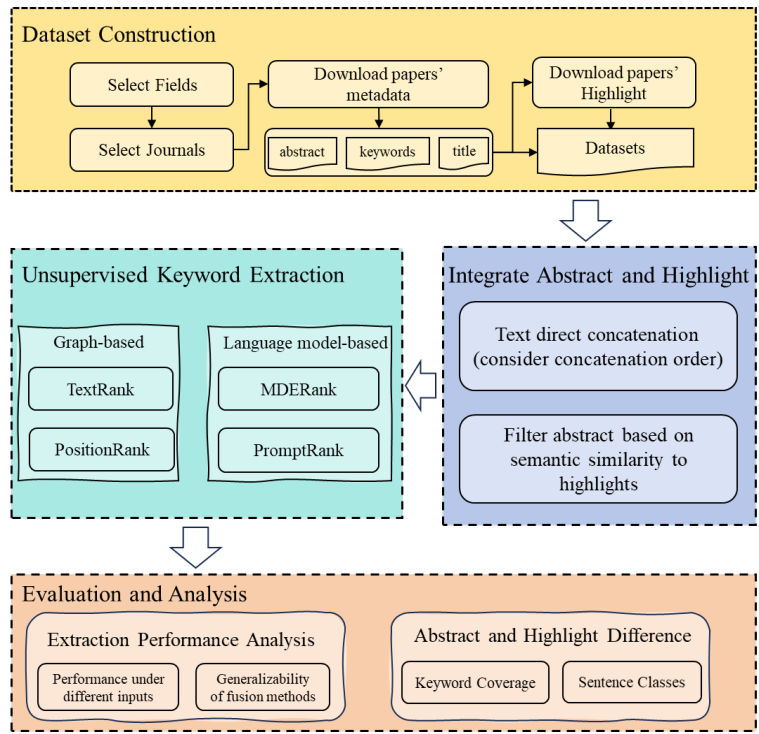

**Fig 2. Framework of this study**

### 3.1 Dataset Construction

Datasets such as Inspec (Hulth, 2003) and SemEval2010 (Kim et al., 2010) are frequently used to evaluate model performance in unsupervised keyword extraction tasks. However, these classic datasets typically contain only abstract and lack the highlights necessary for our study. Therefore, it is essential to construct a new dataset tailored for unsupervised keyword extraction.

After an extensive review of various academic paper sources, we choose Elsevier as the platform for collecting paper highlights for two primary reasons. First, journals published by Elsevier often include highlights in their online versions. Second, the highlights on Elsevier's article detail pages are consistently located, making them amenable to automated extraction through web scraping.

We selected papers from the LIS field and the CS field for analysis, focusing first on identifying representative journals in these two domains, as shown in Table 3. Journals that simultaneously publish research in both domains, such as *Information Processing & Management*, were excluded. In addition, we sought to ensure that the selected journals represent different subfields within each discipline[3]. Since Elsevier journals began to widely adopt highlights after 2012 and our experiment started in May 2024, we collected papers published between 2012 and 2023 from these journals. Based on the Scopus database, we collected the abstract, title, and author-assigned keywords for each paper. The abstracts were used as input for the keyword extraction models, the titles were employed to locate the corresponding articles on the Elsevier website, and the author-assigned keywords served as ground truth for evaluating the performance of the proposed keyword extraction methods.

An initial inspection of the data indicated that CS journals contained significantly more papers than LIS journals. To ensure balance across the datasets, we randomly sampled 100 papers per year from the CS journals, thereby constructing the initial dataset. Subsequently, we counted the number of keywords in each paper and retained only those papers containing between three and six keywords. According to Pottier et al. (2024), most journals set six as the upper limit for keywords, with authors using an average of 5.9 keywords. Excessive keyword counts may introduce redundancy and reduce semantic distinctiveness between keywords. Therefore, this filtering step ensured that the dataset contained papers with a keyword count falling within the typical range. The number of retained papers per journal after this process is shown in the third column of Table 3.

Since the Scopus database does not provide the highlight field, we developed an automated procedure to retrieve it from the online versions of papers. Specifically, we first used the Elsevier API service to obtain the webpage link of each paper based on its title, then analyzed the webpage structure to locate the highlight section. A Python-based web crawler was then implemented to automatically extract highlight content. Some papers were missing highlights in the retrieved data. For these cases, we manually reviewed the records, correcting extraction errors caused by network issues or changes in webpage layout, and removed those papers that did not include highlights at all. The final number of papers per journal is presented in the fourth column of Table 3, where it can be observed that over 80% of papers in each journal contained highlight sections. Finally, we further sampled the CS data to reduce corpus size disparities between the two domains. After all processing steps, the dataset comprised 2,590 papers from the LIS field and 2,996 papers from the CS field. Table 4 provides sample entries from the dataset.

We then analyzed the keyword distribution, abstract sentence length, and highlights sentence length across the two datasets, with the results displayed in Figure 3. Figures 3(a) and 3(d) show that the number of keywords in CS papers typically ranges from 4 to 5, with fewer instances of extremely high or low counts. In contrast, LIS papers predominantly contain 5 keywords, though a significant proportion also have 6 keywords. The distribution of abstract sentence lengths is consistent across both datasets, as illustrated in Figures 3(b) and 3(e). Further analysis revealed that the average sentence length is 8 words in the LIS dataset and 10 words in the CS dataset. This statistic was used in our study to filter out irrelevant information from abstracts. Figures 3(c) and 3(f) depict the average number of highlights sentences per paper in the LIS and CS datasets, respectively, with both averaging around 5 sentences after excluding outliers.

---

[3] Based on manual inspection and expert consultation, we observed that papers in CS field show relatively consistent styles in abstracts, highlights, and keyword usage. Accordingly, we selected four representative subfields familiar to the authors(e.g., Artificial Intelligence, Hardware and Architecture)

Table 3. Selected Journals and the Number of Articles They Contain

| Field | Journal Name | Original Dataset | Final Dataset |
|---|---|---|---|
| LIS | Journal of Informetrics | 955 | 794(83%) |
| | International Journal of Information Management | 1,053 | 881(84%) |
| | Information & Management | 906 | 722(80%) |
| | Library & Information Science Research | 93 | 93 |
| CS | Information Fusion | 1,193 | 1,033(87%) |
| | Journal of Parallel and Distributed Computing | 1,191 | 1,003(84%) |
| | Journal of Systems and Software | 1,195 | 994(83%) |
| | Pattern Recognition | 1,064 | 954(90%) |

Table 4 Data Sample from the Datasets

| Type | Example |
|---|---|
| Keywords | *Collaboration network; Machine learning; Network evolution; Scientific community; SHAP* |
| Abstract | *Scientific communities serve as a fundamental structure of academic activity, and its evolutionary behavior also reveals the development of science…* |
| Highlights | *Interpretable machine learning approaches are applied to the task of event-based Group Evolution Prediction of scientific communities;*<br>*Interpreting the prediction models of community evolution can reflect the underlying mechanisms of the evolution of scientific community;* |

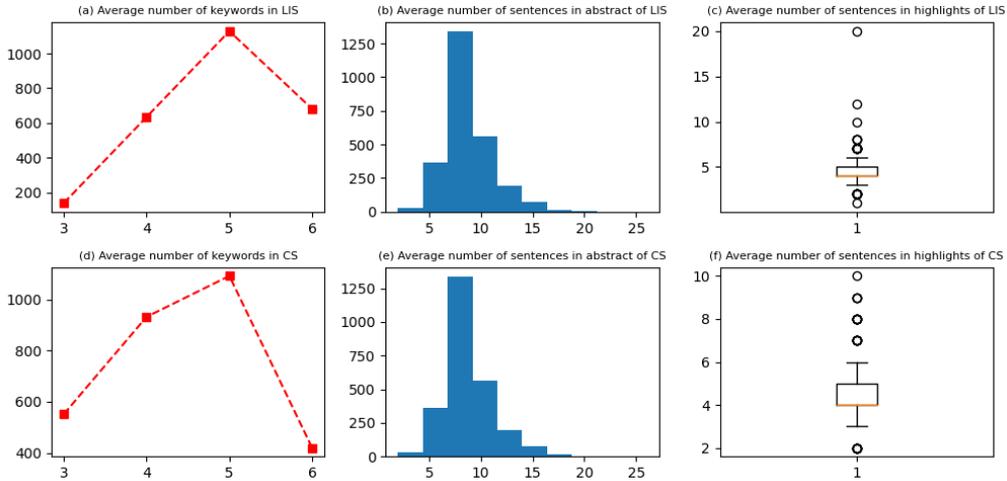

Fig 3. Descriptive Statistics of Datasets

## 3.2 Integrate Abstract and Highlights as Input for Keyword Extraction

Unsupervised keyword extraction involves two critical stages: candidate keyword extraction and candidate keyword scoring. Candidate keywords are typically extracted using rule-based methods, where some of these candidates correspond to actual keywords. The richer the input text in terms of information, the higher the theoretical coverage of keywords within the candidate set. Consequently, integrating highlights information into the abstract can increase the upper limit of keyword extraction task performance. Although the scoring mechanisms for candidate keywords differ slightly depending on the extraction algorithm, there are common patterns. For instance, models often tend to prioritize words that appear earlier in the text as potential keywords, and they typically achieve higher scoring accuracy when the input contains less redundant information. Based on an analysis of the underlying mechanics of unsupervised keyword extraction, this study explores four approaches to integrating highlights and abstract information:

(1) Abstract + Highlights (A+H):

The abstract and highlights texts of the same paper are directly concatenated. In unsupervised keyword extraction, the process typically involves first extracting candidate keywords and then ranking their importance. By concatenating these two types of texts, a larger pool of candidate keywords is generated. If there is overlap between candidate keywords from both texts, the weight of these overlapping keywords increases, resulting in higher importance scores.

(2) Highlights + Abstract (H+A):

The highlights text is placed before the abstract in the concatenation. This approach considers that unsupervised keyword extraction methods often assign higher importance scores to words appearing earlier in the text. While the highlights, being shorter, may not cover keywords as comprehensively as the abstract, the abstract typically contains noise that is irrelevant to keyword extraction. If the abstract is placed at the beginning, non-keyword terms may receive higher scores, which can negatively impact the model's overall performance. By positioning the highlights first, candidate keywords that are more closely aligned with the paper's main themes are given a more advantageous placement.

(3) Filtered Abstract + Highlights (FA+H):

To mitigate the noise present in the abstract, a potential solution is to rank the sentences in the abstract by their semantic similarity to the highlights and retain only the top-k sentences as a filtered abstract. Since the highlights reflects the core contributions and innovations of the paper, it is closely aligned with the paper's main theme. By filtering the abstract in this manner, we are more likely to preserve relevant thematic information while excluding unrelated background content. To implement this, we input the highlights and segmented abstract into a sentence-transformer model to obtain their respective semantic representations. Specifically, we employed *all-MiniLM-L6-v2*, a model trained with contrastive learning that generates a 384-dimensional semantic vector for each input. This model is widely used in tasks such as clustering and semantic search. We then calculate the cosine similarity between the highlights' semantic vector and the semantic vector of each sentence in the abstract, ranking the sentences by similarity, as illustrated in Figure 4. The top-k sentences with the highest similarity scores are selected and recombined to form the filtered abstract.

The value of k plays a crucial role in the outcome: a smaller k retains sentences that are more closely related to the core content of the paper, leading to more accurate importance calculations for candidate keywords, though it may result in an insufficient number of candidate keywords. Conversely, a larger k may introduce more irrelevant information. In this study, we set k to 4 for our experiments, and Section 4.3 analyzes the performance of the extraction method as k varies.

(4) Highlights + Filtered Abstract (H+FA):

The sentence filtering operation for the abstract follows the same procedure as the previous approach, but in this case, the highlights is placed before the filtered abstract.

To evaluate these integrated methods against baseline results, we construct three baselines: using only the abstract (A), only the highlights (H), and only the filtered abstract (FA). We then applied the texts generated by these seven approaches as inputs to the unsupervised keyword extraction methods, assessing their performance using the same metrics. To ensure a fair comparison, we maintain consistent parameter settings across different inputs for the same model, thereby minimizing the influence of the extraction method itself on the results.

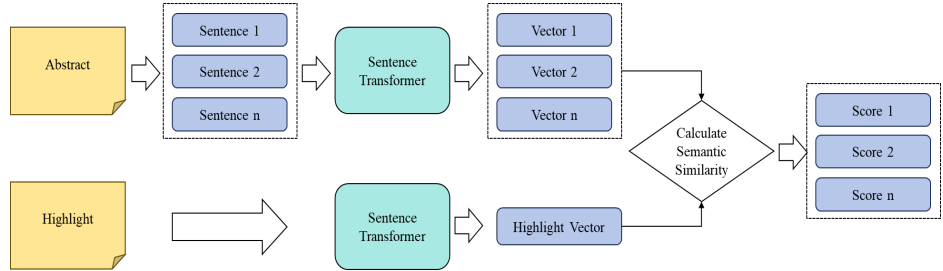

**Fig 4. Abstract noise information filtering based on semantic similarity with highlights**

## 3.3 Model of Unsupervised Keyword Extraction

Unsupervised keyword extraction has been a central topic in computer science. Although the performance of unsupervised methods often trails behind that of supervised models, they offer greater generalizability and are less reliant on extensive training data (Song, et al., 2023a). Moreover, unsupervised keyword extraction evaluates the importance of keywords by analyzing their inherent properties within a document, which provides greater interpretability compared to deep learning models. For these reasons, this study focuses exclusively on unsupervised keyword extraction methods. Following a comprehensive literature review and model comparison, we select the following four models for our experiments:

(1) TextRank

TextRank is a graph-based keyword importance ranking algorithm, conceptually similar to Google's PageRank used for web page ranking (Brin and Page, 1998). In TextRank, words in a document are treated as nodes in a graph, and if two words co-occur within a certain window length, an edge is established between their corresponding nodes. The importance of a node is determined by two principles: nodes with more edges are more important, and if an important node points to another node, that node is also important. The importance of each node in the graph is recursively calculated until convergence, after which the top-k most important words are selected as the keyword set.

(2) PositionRank:

PositionRank incorporates word position and frequency information into the TextRank algorithm. The earlier a word appears in the text and the more frequently it occurs, the higher the weight of its corresponding node in the network. In practice, PositionRank calculates a node's weight as the sum of the reciprocals of all occurrences of a word in the document. The final importance score for each word is then obtained using a recursive calculation similar to that in TextRank.

(3) MDERank:

MDERank is a keyword extraction method based on pre-trained language models. The core idea is that a word that is semantically closer to the abstract is more likely to be a keyword. However, directly calculating the semantic similarity between word vectors and the document vector can lead to mismatches due to differences in text length. To address this, the authors propose masking certain words in the document and then comparing the semantic similarity between the original and masked documents. The higher the similarity, the less important the masked word.

(4) PromptRank:

PromptRank is a keyword extraction method based on an encoder-decoder architecture and prompt learning. The document, from which keywords are to be extracted, is concatenated with a prompt and fed into the encoder. The decoder then calculates the probability of each candidate keyword, with higher probabilities indicating greater importance.

Among these four models, TextRank and PositionRank are graph-based unsupervised keyword extraction models, and our experiments utilize the PyTextRank[4] library to implement them. MDERank[5] and PromptRank[6], on the other hand, are embedding-based methods. We adapt the source code of these models to better fit the dataset used in this study. For PromptRank, we experiment with modifying the prompts in both the encoder and decoder to better align with the input content specific to our experiments. However, these modifications did not result in significant improvements. Consequently, we opt to use the original parameter settings from the published paper.

# 4. Results

In this section, we first introduce the performance evaluation metrics for the models and then analyze the performance of four extraction models using different text combinations as inputs under the LIS and CS datasets. In Section 4.3, we explore the impact of retaining different lengths of abstract content on extraction results when filtering abstract information based on its semantic relevance to the highlights. Section 4.4 focuses on the application of the proposed method to large language models. Finally, we validate the generalizability of the proposed approach on a multidisciplinary corpus.

## 4.1 Evaluation metric

After obtaining the set of keywords extracted by the unsupervised method, we calculate the $F_1$ score between this set and the gold standard keyword set to evaluate the model's performance. The specific calculation method is as follows:

$$P = \frac{len(w_{match})}{len(w_{pred})} \quad (1)$$

$$R = \frac{len(w_{match})}{len(w_{gold})} \quad (2)$$

$$F_1 = \frac{2*P*R}{P+R} \quad (3)$$

Where, $w_{gold}$ and $w_{pred}$ represent the gold standard keyword set and the extracted keyword set, respectively, with $w_{match}$ denoting the intersection of the two sets. In keyword extraction tasks, the $F_1$ scores are typically evaluated at $F_1@5$, $F_1@10$, and $F_1@15$, this corresponds to selecting the top 5, 10, and 15 highest-scoring candidate keywords in $w_{pred}$ during the calculation process. To achieve more accurate model evaluation results, we filtered out duplicate keywords in $w_{pred}$ before computing the metric scores and performed stemming on the keywords in both $w_{gold}$ and

---

[4] https://pypi.org/project/pytextrank/
[5] https://github.com/LinhanZ/mderank
[6] https://github.com/NKU-HLT/PromptRank

$w_{pred}$ sets.

We further analyze the statistical significance of performance improvements for other input conditions compared to the baseline condition, which uses only the Abstract as input. Specifically, for each document, we calculated the keyword extraction performance score. We then employed a paired t-test to compare the scores obtained under an experimental input condition with the scores obtained under the baseline condition (Abstract-only). Performance improvement is considered statistically significant if the t-test shows a significant difference between the two groups of scores and the experimental input condition yields superior results to the Abstract-only input. To rigorously control the probability of Type I errors, we established three standard alpha ($\alpha$) levels for our significance testing: $\alpha = 0.1$, $\alpha = 0.05$, and $\alpha = 0.01$. This configuration is also a common standard in information retrieval and natural language processing research (Smucker et al., 2007). The rationale for using these multiple thresholds is to provide a comprehensive evaluation of performance variations across different input settings: specifically, $\alpha = 0.1$ enables us to capture marginal but meaningful improvements, whereas $\alpha = 0.01$ clearly highlights the settings that yield the most pronounced and robust gains. These significance levels are denoted by *, **, and *** respectively in the subsequent tables.

## 4.2 Extraction Performance Based on Abstract and Highlights

Tables 5 and 6 present the keyword extraction performance for the LIS and CS datasets under different input conditions. From a model perspective, keyword extraction methods based on pre-trained language models consistently outperform graph-based methods. For instance, in the LIS dataset, the $F_1@5$ scores for TextRank and PositionRank are both below 18%, whereas MDERank and PromptRank achieve around 22%. Although the performance gap between these two types of models is less pronounced in the CS dataset, MDERank and PromptRank still exhibit superior results. This underscores the advantage of leveraging semantic relationships between candidate words and the text, highlighting the robust capabilities of pre-trained language models in handling various downstream tasks.

When comparing the performance across different inputs, models that combine highlights with abstract outperform those that rely on a single source of information. This improvement is particularly noticeable in keyword extraction methods based on pre-trained language models. For example, with the MDERank model on the LIS dataset, the $F_1@5$ score is 19.48% when using only the abstract, 16.97% when using only highlights, and increases to 22.2% when both are concatenated. On the CS dataset, the extraction performance is 12.02% using the abstract alone, 8.95% using only the highlights, and improves to 13.88% when both are combined. These findings validate the effectiveness of the proposed method, demonstrating that integrating highlights with abstract can significantly enhance the performance of traditional unsupervised keyword extraction tasks.

When comparing different integration methods, we find that altering the concatenation order of highlights and abstract does not significantly affect model performance. In experiments involving filtered abstract, as reflected in Tables 5 and 6, there is no substantial difference in metric scores between using the full abstract and the filtered abstract across various models. This suggests that in unsupervised keyword extraction experiments, only a small portion of the abstract is valuable to the model. To further validate this, we used the average number of sentences per abstract in each dataset as a threshold and compared performance when retaining different numbers of sentences. Table 7 presents the results for *k* = 4 versus the best-performing *k* value, showing that retaining only 4–5 sentences is generally sufficient to achieve strong keyword extraction performance. This finding not only reinforces the above conclusion but also indirectly supports the effectiveness of our proposed highlight-based filtering method for extracting core information from abstracts.

Comparing the experimental results between the two datasets, we observe that the LIS domain achieves better keyword extraction performance. The highest $F_1@5$ score on the LIS dataset is 23.71%, compared to only 16.91% on the CS dataset. Since all keyword extraction methods used in this study are unsupervised, and the descriptive statistics in Figure 3 show no significant differences between the two datasets in terms of keyword count, text length, or other factors, we can rule out the influence of the model or data scale. One possible explanation for this disparity is that scholars in the LIS field tend to select keywords directly from important words within the paper, whereas scholars in the CS field often derive keywords from a summary of the paper's content—these keywords may not necessarily appear in the original text. Consequently, extraction-based methods may struggle to accurately extract keywords from CS abstract. In Section 5.1, we conduct a statistical analysis of keyword coverage in the abstract of both fields to further explore this issue.

**Table 5. Performance of keyword extraction on the LIS corpus across various input conditions**

| $F_1@K$ | Method | Input | | | | | | |
|---|---|---|---|---|---|---|---|---|
| | | A | H | FA | A+H | H+A | FA+H | H+FA |
| 5 | TextRank | **17.79** | 12.09 | 17.71 | 17.77 | 17.69 | 17.53 | 17.60 |
| | PositionRank | 17.68 | 12.40 | 17.47 | 17.64 | **18.02** | 17.55 | 17.63 |
| | MDERank | 19.48 | 16.97 | 18.35 | **22.32**[***] | 22.20[***] | 21.37[**] | 21.41[*] |
| | PromptRank | 22.33 | 16.14 | 21.92 | 23.06[***] | **23.71**[***] | 22.56 | 22.49 |
| 10 | TextRank | 17.66 | 10.82 | 16.63 | 18.01[**] | **18.02**[***] | 17.47 | 17.51 |
| | PositionRank | 17.06 | 10.78 | 16.36 | 17.58[***] | **18.24**[***] | 17.19 | 17.28 |
| | MDERank | 18.91 | 15.70 | 18.66 | 21.36[***] | **21.37**[***] | 20.69 | 20.61 |
| | PromptRank | 21.72 | 13.78 | 19.75 | 22.78[***] | **23.08**[***] | 21.73 | 21.68 |
| 15 | TextRank | 15.56 | 9.49 | 13.95 | 16.29[***] | **16.45**[***] | 15.35 | 15.31 |
| | PositionRank | 15.26 | 9.50 | 13.71 | 15.97[***] | **16.41**[***] | 15.19 | 15.16 |
| | MDERank | 17.17 | 13.27 | 16.89 | 19.21[***] | **19.27**[***] | 18.49[**] | 18.52[**] |
| | PromptRank | 18.78 | 12.76 | 16.37 | 19.95[***] | **20.28**[***] | 18.39 | 18.34 |

*Note: Each row reports the overall extraction performance of a given keyword extraction model under a specific input setting. Bold values indicate the best performance in each row. Significance markers denote improvements over using only the abstract: \*\*\*p < 0.01, \*\*p < 0.05, \*p < 0.1. Same below.*

**Table 6. Performance of keyword extraction on the CS corpus across various input conditions**

| $F_1@K$ | Method | Input | | | | | | |
|---|---|---|---|---|---|---|---|---|
| | | A | H | FA | A+H | H+A | FA+H | H+FA |
| 5 | TextRank | 11.76 | 6.69 | 10.98 | **12.22**[***] | **12.22**[***] | 11.31 | 11.40 |
| | PositionRank | 12.52 | 6.96 | 10.55 | **12.89**[***] | 12.70 | 11.32 | 11.45 |
| | MDERank | 12.02 | 8.95 | 11.11 | **14.02**[***] | 13.99[***] | 13.07[***] | 12.88[***] |
| | PromptRank | 16.44 | 8.44 | 14.24 | **16.91**[***] | 16.53 | 14.84 | 14.53 |
| 10 | TextRank | 11.43 | 5.64 | 9.74 | 11.98[***] | **12.03**[***] | 10.90 | 10.92 |
| | PositionRank | 11.46 | 5.66 | 9.48 | **12.19**[***] | 12.12[***] | 10.64 | 10.64 |
| | MDERank | 12.18 | 8.06 | 11.49 | 13.88[***] | **13.95**[***] | 12.76[***] | 12.91[***] |
| | PromptRank | 15.03 | 7.50 | 12.31 | **15.89**[***] | 15.82[***] | 13.80 | 13.64 |
| 15 | TextRank | 9.99 | 5.04 | 7.95 | 10.64[***] | **10.81**[***] | 9.18 | 9.18 |
| | PositionRank | 9.94 | 5.02 | 7.84 | 10.70[***] | **10.74**[***] | 9.10 | 9.10 |
| | MDERank | 11.54 | 7.06 | 10.58 | 12.80[***] | **12.88**[***] | 11.71[**] | 11.69[***] |
| | PromptRank | 12.95 | 7.19 | 10.42 | **13.90**[***] | 13.85[***] | 11.69 | 11.60 |

Table 7. Model performance under different numbers of retained abstract sentences

| $F_1@K$ | Method | LIS | | CS | |
|---|---|---|---|---|---|
| | | k=4 | best_k | k=4 | best_k |
| 5 | TextRank | 17.71 | -- | 10.98 | 10.99(6) |
| | PositionRank | 17.47 | -- | 10.55 | 10.90(7) |
| | MDERank | 18.35 | 18.49(5) | 11.11 | -- |
| | PromptRank | 21.92 | 22.17(5) | 14.24 | 14.61(5) |
| 10 | TextRank | 16.63 | 16.89(5) | 9.74 | 10.47(6) |
| | PositionRank | 16.36 | 16.61(5) | 9.48 | 10.23(6) |
| | MDERank | 18.66 | -- | 11.49 | 11.70(5) |
| | PromptRank | 19.75 | 20.98(6) | 12.31 | 13.53(7) |
| 15 | TextRank | 13.95 | 14.80(6) | 7.95 | 8.98(7) |
| | PositionRank | 13.71 | 14.65(6) | 7.84 | 8.87(7) |
| | MDERank | 16.89 | 17.15(5) | 10.58 | 11.01(6) |
| | PromptRank | 16.37 | 17.87(7) | 10.42 | 11.70(8) |

*Note: The best_k column reports the highest metric score achieved for each model and metric under different numbers of retained sentences. The value in parentheses indicates the corresponding number of retained sentences. Empty cells indicate that the best score was obtained under the default setting (k = 4).*

## 4.3 The Impact of Sentence Count in Filtered Abstract on Keyword Extraction Results

Section 4.2 presents experimental results using filtered abstract as model input, demonstrating that retaining only a portion of the abstract can yield keyword extraction performance comparable to using the entire abstract. This finding prompts the question of the optimal number of sentences that should be retained in the filtered abstract to achieve the best extraction results. This section investigates how keyword extraction performance varies with different numbers of retained sentences in the filtered abstract. Specifically, we set the average number of sentences in abstract as the upper limit for this analysis. By experimenting with filtered abstract that contain varying numbers of sentences, we aim to provide a more intuitive understanding of the proportion of valuable information within abstract and its impact on extraction performance.

In Section 3.1, we conduct a statistical analysis of the sentence distribution in the abstract of the two datasets, finding that the average number of sentences in LIS dataset abstract is 8, while in the CS dataset, it is 10. For consistency in our analysis, we set the maximum number of sentences in the filtered abstract to 8. Using the method proposed in Section 3.3, we ranked the importance of sentences in the original abstract based on their semantic similarity to the highlights and extracted the top K sentences with the highest semantic similarity to form the filtered abstract, with K ranging from 1 to 8. For records where the number of sentences in the original abstract is less than K, the sentences are reassembled based on semantic similarity from high to low to obtain the "filtered" abstract.

Figures 5 and 6 illustrate how the performance of four models changes as the number of sentences in the filtered abstract varies in the LIS and CS datasets. Taking Figure 5 as an example, different colored lines in each subfigure represent different models. The solid lines indicate the metric scores when using filtered abstract of varying lengths as input, while the dashed lines indicate the extraction performance when using the full abstract as input. Subfigures (a), (b), and (c) correspond to $F_1@5$, $F_1@10$, and $F_1@15$ metric scores, respectively. The overall trend observed indicates a performance pattern that initially increases with the number of sentences in the input before declining. We hypothesize that while adding more text provides additional candidate keywords, thereby enhancing the potential for higher extraction performance, it also introduces noise that complicates keyword identification. Consequently, both an insufficient and an excessive number of sentences in the input can negatively impact extraction performance.

Moreover, different evaluation metrics exhibit varying sensitivity to the number of sentences. For instance, in the best-performing PromptRank model, the optimal number of sentences is 5 for the $F_1@5$ metric, whereas the optimal numbers for the $F_1@10$ and $F_1@15$ metrics are 6 and 7, respectively. Although this pattern varies slightly across other models, the general trend remains consistent. This variation likely stems from the calculation method of the $F_1@k$ metric, which compares the top k predicted keywords with the top k actual keywords. As mentioned earlier, increasing the sentence length has dual effects: it expands the candidate keyword pool but also adds more noise. When only the top 5 keywords are compared, the model can achieve higher accuracy with a lower coverage requirement for candidate keywords. However, when the top 10 or 15 keywords are considered, a broader range of candidate keywords must be covered, leading to longer optimal sentence lengths for these metrics compared to $F_1@5$.

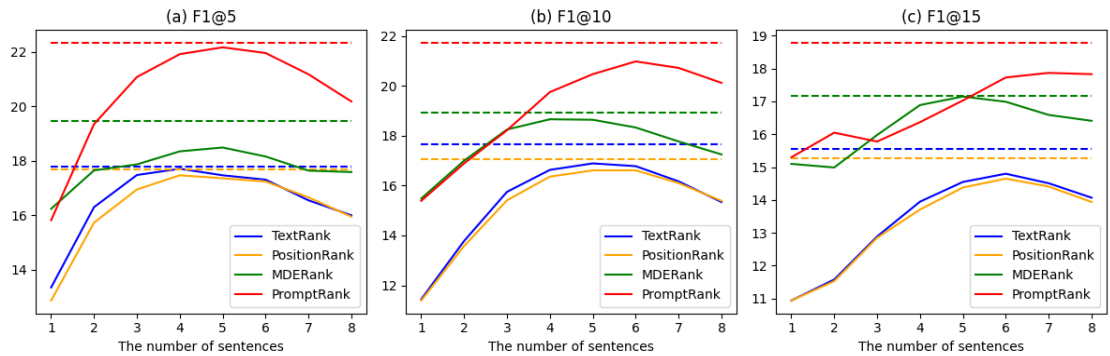

*Note: The solid line in the figure represents the model's extraction performance when retaining different numbers of sentences in the abstract as input. The dotted line indicates the model's extraction performance when using the complete abstract as input. Same below.*

**Fig 5. The law of performance change with the number of abstract sentences on the LIS dataset**

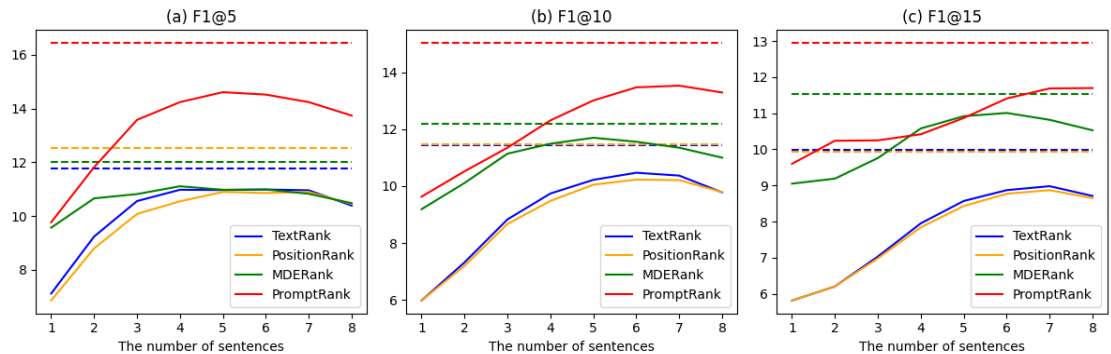

**Fig 6. The law of performance change with the number of abstract sentences on the CS dataset**

## 4.4 The Impact of Integrate Methods on Keyword Extraction Using Large Language Models

To evaluate the generalizability of our proposed method across different architectures, we conducted experiments using GPT-4o (gpt-4o-mini-2024-07-18). Furthermore, acknowledging that Large Language Model (LLM) performance is often sensitive to prompt design and that proprietary models may exhibit temporal instability—as previous studies have reported performance fluctuations over time (Chen et al., 2023; Chen et al., 2024)—we also employed the open-source model LLaMA-3 (llama-3-8B-Instruct).

For proprietary (non-open-source) models, we accessed it via API calls. Specifically, we first designed the following prompt template: *"Based on the text below, give the keyword set of the paper in which it is located, sort the results from high to low importance, and separate the results with commas. Note that only the results are given without any additional explanations."* The abstract, highlights, and other textual content were then appended to this prompt as input. Depending on the model type, the corresponding API key was retrieved, and all calls were implemented in a unified manner through the LangChain framework. For parameters affecting output quality, such as *top_p* and *max_tokens*, we adopted LangChain's default settings. Specifically, *top_p* was set to 1 to allow maximum creativity, while *max_tokens* was left unrestricted. For the open-source model LLaMA-3, we deployed it locally in an environment equipped with two A40 GPUs. Different inputs were subsequently fed into the model to obtain generated results. The design of the prompt template and model generation parameters (e.g., *top_p*, *temperature*) was kept consistent with that used for the proprietary models to ensure comparability.

Tables 8 and 9 summarize the model performance on the LIS and CS datasets. Aside from GPT-4o on the CS dataset, incorporating highlights into abstracts yielded consistent and significant performance gains across all other configurations. This attests to the robustness of our approach: irrespective of the backbone model, fusing highlights with abstracts outperforms traditional abstract-only methods in unsupervised keyword extraction. In terms of model comparison, LLaMA-3 largely surpassed GPT-4o, barring specific inputs limited to highlights or filtered abstracts. This highlights the viability of open-source models, which provide a cost-effective alternative to proprietary, token-billed APIs without compromising on performance.

Table 8. Comparative Performance of Large Language Models on the LIS Dataset

| $F_1@K$ | Method | Input | | | | | | |
|---|---|---|---|---|---|---|---|---|
| | | A | H | FA | A+H | H+A | FA+H | H+FA |
| 5 | GPT-4o | 28.35 | 24.87 | 27.44 | 30.03[***] | **30.30**[***] | 28.18 | 28.09 |
| | llama-3 | 28.59 | 21.46 | 24.17 | **30.72**[***] | 30.55[***] | 27.72 | 26.85 |
| 10 | GPT-4o | 25.67 | 22.62 | 23.83 | 26.07[*] | **26.20**[*] | 24.36 | 24.39 |
| | llama-3 | 25.60 | 20.27 | 21.73 | **26.89**[***] | 26.76[***] | 24.75 | 24.14 |
| 15 | GPT-4o | 22.63 | 22.20 | **22.91**[*] | 22.89[*] | 22.85 | 22.68[*] | 22.72 |
| | llama-3 | 23.83 | 20.00 | 21.12 | **24.62**[***] | 24.46[**] | 23.38 | 22.87 |

Table 9. Comparative Performance of Large Language Models on the CS Dataset

| $F_1@K$ | Method | Input | | | | | | |
|---|---|---|---|---|---|---|---|---|
| | | A | H | FA | A+H | H+A | FA+H | H+FA |
| 5 | GPT-4o | 24.82 | 17.25 | 20.91 | **24.91** | 23.49 | 21.08 | 20.73 |
| | llama-3 | 25.00 | 15.73 | 19.61 | **26.96**[***] | 26.20[***] | 22.77 | 22.77 |
| 10 | GPT-4o | **21.05** | 14.99 | 17.67 | 20.51 | 19.67 | 17.54 | 17.55 |
| | llama-3 | 21.56 | 14.53 | 17.34 | **22.81**[***] | 22.39[***] | 19.56 | 19.56 |
| 15 | GPT-4o | **18.21** | 14.71 | 16.90 | 17.62 | 17.04 | 16.11 | 16.25 |
| | llama-3 | 19.83 | 14.40 | 16.82 | **20.55**[**] | 20.29 | 18.43 | 18.44 |

## 4.5 Robustness check

We evaluated the role of highlights in unsupervised keyword extraction across research articles from multiple disciplinary domains. Following the ASJC classification of 27 subject areas, we randomly selected one sub-discipline from each field, then chose a representative journal under Elsevier, and further sampled 20 research articles from each journal. This process yielded a multidisciplinary experimental corpus comprising 540 research articles. The selected journals and their corresponding subject areas are listed in Table 10. Due to variations in publication requirements across journals, the dataset contains abstracts of different formats (i.e., traditional abstracts and structured abstracts). In addition, notable differences exist in the number of author-assigned keywords (e.g., a larger number in Neuroscience and Immunology & Microbiology), the descriptive style of highlights, and the length of highlights across papers. These variations enable the dataset to better demonstrate the applicability of our proposed method across diverse textual corpora, especially given prior findings that keyword distributions (Gupta et al., 2025) and writing style of academic papers (Dong et al., 2024) differ across disciplines and databases.

Table 10. Journal Composition of the Multidisciplinary Experimental Corpus

| No | Field | Journal Name |
|---|---|---|
| 1 | Agricultural And Biological Sciences | Ecosystem Services |
| 2 | Arts and Humanities | Journal of Aging Studies |
| 3 | Biochemistry, Genetics and Molecular Biology | Cell |
| 4 | Business, Management and Accounting | Journal of Economics and Business |
| 5 | Chemical Engineering | Chemical Engineering Journal |
| 6 | Chemistry | Carbon |
| 7 | Computer Science | Pattern Recognition |
| 8 | Decision Sciences | Ecological Indicators |
| 9 | Dentistry | Dental Materials |
| 10 | Earth and Planetary Sciences | Advanced in Space Research |
| 11 | Economic, Econometrics and Finance | Journal of Financial Stability |
| 12 | Energy | Applied Energy |
| 13 | Engineering | Composites Science and Technology |
| 14 | Environmental Science | Journal of Cleaner Production |
| 15 | Health Professions | Human Factors in Healthcare |
| 16 | Immunology and Microbiology | Journal of Theoretical Biology |
| 17 | Materials Science | Energy Storage Materials |
| 18 | Mathematics | Applied Mathematical Modelling |
| 19 | Medicine | American Journal of Emergency Medicine |
| 20 | Multidisciplinary | Journal of Advanced Research |
| 21 | Neuroscience | Brain behavior and Immunity |
| 22 | Nursing | Complementary Therapies in Medicine |
| 23 | Pharmacology, Toxicology and Pharmaceutics | Life Sciences |
| 24 | Physics and Astronomy | European Polymer Journal |
| 25 | Psychology | Journal of Vocational Behavior |
| 26 | Social Sciences | Social Networks |
| 27 | Veterinary | Veterinary Microbiology |

The results of unsupervised keyword extraction on this dataset are presented in Table 11. Even in a multidisciplinary setting, incorporating highlights into the abstract consistently improves extraction performance. Notably, in terms of $F_1@5$ and $F_1@10$, models using only highlights outperform those using only abstracts. Furthermore, the combination of filtered abstracts with highlights achieves the best extraction performance under the $F_1@5$ metric. This finding further indicates that abstracts often contain redundant information, and that highlights serve as an effective mechanism both to filter the abstract and to provide complementary information, thereby enhancing the precision and coverage of keyword extraction. Overall, these results suggest that despite disciplinary, editorial, and stylistic differences across research articles, highlights consistently provide valuable contextual information for unsupervised keyword extraction.

Table 11. Results of Unsupervised Keyword Extraction on the Multidisciplinary Corpus

| $F_1@K$ | Method | Input | | | | | | |
|---|---|---|---|---|---|---|---|---|
| | | A | H | FA | A+H | H+A | FA+H | H+FA |
| 5 | TextRank | 9.55 | 9.89 | 9.55 | 9.99 | 9.95 | 10.51** | **10.62**** |
| | PositionRank | 9.74 | 10.44 | 9.85 | 10.59** | 10.07 | **11.42**** | 10.70 |
| | MDERank | 11.87 | 11.99 | 11.42 | 14.09*** | **14.17**** | 13.46*** | 13.17** |
| | PromptRank | 13.57 | 11.25 | 11.72 | 14.65*** | **14.95**** | 13.80** | 13.57 |
| 10 | TextRank | 9.99 | 10.06 | 9.73 | **11.50**** | 11.21*** | 11.35*** | 11.38*** |
| | PositionRank | 10.22 | 10.08 | 9.68 | 11.38*** | **11.63**** | 11.30** | 11.60*** |
| | MDERank | 12.00 | 11.83 | 11.73 | **13.76**** | **13.76**** | 13.51*** | 13.63*** |
| | PromptRank | 13.41 | 10.45 | 11.46 | 14.75*** | **14.82**** | 14.45** | 13.93 |
| 15 | TextRank | 9.18 | 8.97 | 8.44 | **10.74**** | 10.72*** | 10.70*** | 10.69*** |
| | PositionRank | 9.52 | 8.97 | 8.57 | 10.90*** | **11.27**** | 10.52*** | 10.78*** |
| | MDERank | 11.18 | 10.90 | 10.81 | 12.87*** | **12.96**** | 12.76*** | 12.48*** |
| | PromptRank | 11.79 | 9.80 | 9.78 | **13.53**** | 13.53*** | 12.69** | 12.71*** |

In addition to validating the generalizability of our approach across different subject domains, we further extended our experiments to journals outside Elsevier that also mandate the inclusion of highlights. Specifically, we selected *Environmental Processes* under Springer as a target journal. Unlike Elsevier journals, where highlights appear only in the online version, *Environmental Processes* requires authors to include highlights alongside the abstract and keywords within the main text, reflecting the journal's emphasis on this structural component. We randomly sampled 100 articles from this journal and extracted abstracts, highlights, and keywords from the PDF files for evaluation. The experimental results are shown in Table 12. Across all models and evaluation metrics, incorporating highlights—either by directly appending them to the abstract or by combining them with filtered abstracts—consistently and significantly improved performance compared to using abstracts alone. For example, under the $F_1@5$ metric, the PromptRank model improved from 11.40 to 14.02, while TextRank increased from 8.79 to 12.15 after incorporating highlights. These findings provide further evidence for the generalizability of our conclusions, demonstrating that the performance benefits of integrating highlights into unsupervised keyword extraction are not constrained by specific research domains, journals, or publishers' editorial requirements.

Table 12． Results of Unsupervised Keyword Extraction on the Non-Elsevier Journal

| $F_1@K$ | Method | Input | | | | | | |
|---|---|---|---|---|---|---|---|---|
| | | A | H | FA | A+H | H+A | FA+H | H+FA |
| 5 | TextRank | 8.79 | 10.84 | 8.60 | 11.59*** | 11.40*** | **12.15**** | **12.15**** |
| | PositionRank | 9.72 | 10.46 | 10.09 | 11.21* | 12.15 | 11.78 | **12.34**** |
| | MDERank | 9.71 | 11.40 | 9.72 | **12.71**** | 12.52*** | 11.96 | 12.71** |
| | PromptRank | 11.40 | 11.67 | 11.40 | 12.90* | **14.02**** | 12.52 | 12.90 |
| 10 | TextRank | 9.55 | 11.00 | 9.43 | **11.97**** | 11.97*** | 11.97** | 11.97** |
| | PositionRank | 9.68 | 10.61 | 10.70 | 12.10*** | 12.48** | **13.25**** | 12.61** |
| | MDERank | 12.35 | 12.00 | 12.29 | **14.78**** | 14.65*** | 14.27* | 13.12 |
| | PromptRank | 12.99 | 10.89 | 12.49 | 14.90*** | **16.18**** | 14.90* | 14.78* |
| 15 | TextRank | 9.18 | 9.71 | 9.95 | 11.59*** | 11.59*** | **12.27**** | 11.98*** |
| | PositionRank | 10.05 | 9.71 | 10.05 | 11.50* | 11.88** | **12.37**** | 11.88* |
| | MDERank | 11.78 | 11.04 | 11.50 | 14.20*** | **14.40**** | 13.53* | 13.24 |
| | PromptRank | 12.66 | 10.52 | 11.24 | **15.17**** | 14.78*** | 13.82 | 13.53 |

# 5. Mechanism Analysis of Highlights-Driven Enhancement in Performance of Unsupervised Keyword Extraction

The integration of highlights and abstract in unsupervised keyword extraction demonstrates improved performance compared to using abstract alone. While previous sections provide a qualitative explanation for this improvement, attributing it to the enriched information in the input, this section aims to offer a more precise, quantitative analysis. We investigate how the coverage of keywords within the text under different inputs correlates with model performance. Additionally, we examine the differences in textual content between abstract and highlights to better understand the strengths and weaknesses of keyword extraction based on each text type. This analysis helps elucidate the mechanisms behind the observed performance enhancements.

## 5.1 Differences in Keyword Coverage Across Different Input Texts

In unsupervised keyword extraction, the first crucial step is selecting candidate keywords from the text. The quantity and quality of these candidate keywords set the upper limit on the extraction model's performance. In this section, we focus on extracting candidate keywords from three types of texts: abstract, highlights, and abstract with added highlights. We then calculate the proportion of true keywords within the candidate keyword set, which we refer to as keyword coverage. By analyzing the differences in keyword coverage among these three types of texts, we aim to explore the potential reasons for the varying extraction performance observed with different inputs.

For candidate keyword extraction, we use the StanfordNLP[7] and NLTK[8] libraries, both of which are Python-based natural language processing tools capable of performing tasks such as sentence splitting, word segmentation, and part-of-speech tagging. Following established practices in unsupervised extraction model development, we first use StanfordNLP to obtain the part-of-speech tags for terms in the input text. The tagged results are then input into the NLTK library to parse into a syntactic tree. We traverse this tree and retain terms with a part-of-speech combination of adjective + noun or nouns only as candidate keywords.

We compare these candidate keywords with the gold standard keywords, calculating the keyword coverage for different input texts across two datasets, as shown in Tables 13 and 14. Each column in the tables indicates the number of papers where *k* keywords are present in the candidate keyword set for the given input text. The "A+H" column represents the combined input of abstract and highlights, "A-H" represents papers where keywords appear in the abstract but not in the highlights, and "H-A" represents papers where keywords appear only in the highlights.

From the "H-A" column, we observe that about 450 papers in both datasets have keywords that appear exclusively in the highlights, accounting for 17.6% and 14.6% of the total papers in each dataset, respectively. This finding suggests that, for these papers, integrating highlights into the abstract increases the keyword coverage in the input text, thereby raising the potential upper limit of extraction performance. Further analysis of the "A" and "A+H" columns reveals that combining highlights with the abstract increases the number of samples containing three or more keywords. Given that this study controlled for the number of keywords per paper to be between three and six, this fused input is likely to theoretically improve recall rates, thereby enhancing

---

[7] https://stanfordnlp.github.io/stanfordnlp/
[8] https://www.nltk.org/

overall performance.

Additionally, we observe some interesting phenomena. The keyword coverage in highlights alone is significantly lower than in the abstract, as reflected in the experimental results in Tables 5 and 6. This lower coverage suggests that using only highlights as input leads to poorer extraction performance, likely because the shorter length of highlights cannot encompass a broader range of content information. While this section provides an initial explanation of how keyword coverage in the input text influences extraction performance after integrating highlights, it's important to note that unsupervised extraction methods also involve ranking the importance of candidate keywords. Therefore, we cannot guarantee that the keywords provided by highlights will be recognized and output by the model. This issue is complex and will be addressed in future research.

Another phenomenon highlighted by Tables 13 and 14 is that keywords in computer science (CS) papers are less frequently present in the content. In the LIS dataset, the proportions of papers without any keywords in the A, H, and A+H approaches are 9%, 29%, and 7%, respectively. In contrast, in the CS dataset, these proportions are 18%, 56%, and 14%, respectively. This disparity suggests that authors in the CS domain tend to provide keywords that are more abstracted from the original content, meaning these terms may not necessarily appear in the paper. This finding aligns with the experimental results in Tables 5 and 6, where extraction performance on the LIS dataset is consistently higher than that on the CS dataset under the same model and input conditions.

Table 13. Keyword coverage of various types of text in the LIS dataset

| Number of Gold Standards | Input | | | | |
|---|---|---|---|---|---|
| | A | H | A+H | A-H | H-A |
| 0 | 240 | 742 | 181 | 703 | 2,134 |
| 1 | 546 | 929 | 468 | 958 | 397 |
| 2 | 738 | 600 | 705 | 625 | 54 |
| 3 | 591 | 243 | 663 | 233 | 4 |
| 4 | 334 | 63 | 384 | 61 | 0 |
| 5 | 124 | 12 | 167 | 9 | 0 |
| 6 | 16 | 0 | 21 | 0 | 0 |

Table 14 Keyword coverage of various types of text in the CS dataset

| Number of Gold Standards | Input | | | | |
|---|---|---|---|---|---|
| | A | H | A+H | A-H | H-A |
| 0 | 531 | 1,663 | 417 | 873 | 2,559 |
| 1 | 1,028 | 966 | 972 | 1,193 | 390 |
| 2 | 837 | 296 | 876 | 648 | 45 |
| 3 | 431 | 62 | 504 | 218 | 2 |
| 4 | 142 | 9 | 187 | 55 | 0 |
| 5 | 24 | 0 | 37 | 9 | 0 |
| 6 | 3 | 0 | 3 | 0 | 0 |

## 5.2 Differences in Content Between Abstract and Highlights

The differences between abstract and highlights are immediately apparent in terms of both length and content. Abstract typically provide a detailed description of the background, methodology, and conclusions, resulting in more sentences that are relatively longer. In contrast, highlights generally consist of about four concise sentences designed to leave a strong impression on the reader regarding the paper's innovations and contributions. This disparity also extends to keyword coverage, as discussed in detail in Section 5.1.

In this section, we examine the differences in focus between abstract and highlights. To facilitate a direct comparison, we apply text classification methods to identify the distribution of sentence types within each text. The text classification experiment involves two critical aspects: selecting a classification system and determining a classification method. For the former, we find that the academic community has developed a relatively comprehensive abstract classification system, such as the one used in structured abstract studies, which divides abstract into sections like background, methods, and conclusions. However, these studies typically classify blocks of sentences rather than individual sentences, making the models trained on this data less effective at accurately identifying the types of individual sentences in highlights. For classifying highlights, Wang et al. (2023) proposed the system shown in Table 14, which includes five categories: research introduction, purpose and background, process and methods, conclusion, and contributions and promotion. While Wang did not apply this system to abstract sentences, our experience suggests that these five categories can also cover most of the content found in abstract. Therefore, we based our experiments on the classification schema outlined in Table 15.

Table 15. Sentence category schema for abstract and highlights

| Category | Description | Example |
| --- | --- | --- |
| Research Introduction | An overview of the research subject, research question, or research outcome | *A study of attitudes to geolocational data harvesting via smartphones* |
| Purpose and Background | The reason for or meaning of the research and an associated explanation | *Current optimizations do not fully utilize the memory bandwidth and computing power* |
| Process and Methods | How the research is implemented | *We evaluate the algorithm on software with complex input domains* |
| Results or Conclusion | Different types of research outcomes include theories, phenomena, patterns, mechanisms, techniques, and relationships | *The joint bending stiffness has a great influence on the buckling load capacity of latticed shell structures* |
| Contribution and Promotion | Author-evaluated contribution of the research | *Proper fusion of deep learning methods outperforms the state-of-the-art* |

Regarding the classification model, since Wang's study did not publicly release its dataset, we employed GPT-4o (*gpt-4o-mini-2024-07-18)* for our experiments. This choice enabled us to focus on result analysis rather than manual sentence categorization. Specifically, we incorporated the 13 annotated samples provided by Wang into the prompt as exemplars, supplemented with task descriptions and category information. The prompt, together with the target sentences, was then submitted to GPT-4o via API calls to obtain classification outputs. The model invocation procedure was kept consistent with that described in Section 4.4.

Figure 7 illustrates the distribution of sentence categories in abstract and highlights for the LIS and CS datasets. The bar graphs in each subplot depict the number of sentences corresponding to each category, while the line graphs represent the proportion of specific sentence types. In the LIS dataset, a higher proportion of sentences are dedicated to conclusions or summaries and research introductions, with relatively fewer sentences addressing research motivation, methods, and contributions. The most notable difference between highlights and abstract in the LIS dataset is found in the conclusions or summaries. Conversely, the CS dataset shows a different pattern, with sentences describing the research process and methods dominating both abstract and highlights. The primary distinction between abstract and highlights in the CS dataset lies in the purpose and background, and contributions and advocacy sentence types; the former is more prominent in abstract, while the latter is more prevalent in highlights.

The differences in distribution across these two datasets can be attributed to the distinct scientific paradigms and research methods of various disciplines. In LIS, researchers may use bibliometric analyses or machine learning techniques, but they often place greater emphasis on interpreting the results. In contrast, CS scholars prioritize designing new algorithms to improve the performance of existing solutions. This divergence results in a greater prevalence of results or conclusion sentences in LIS, while process and method sentences are more common in CS. The observed differences in sentence types between abstract and highlights within each dataset suggest that the motivations behind writing these two types of text are distinct. Highlights are not merely condensed versions of abstract; instead, authors tend to emphasize the conclusions and contributions of their research in highlights, allowing readers to quickly grasp the paper's value.

To further understand how differences in content between abstract and highlights influence the performance of unsupervised extraction models, we analyze the keyword coverage of each sentence type, as depicted in Figures 7 and 8. Initially, we hypothesize that conclusion sentences in the LIS dataset and method sentences in the CS dataset will exhibit the highest keyword coverage. Surprisingly, however, it is the research introduction sentences—regardless of whether they are from abstract or highlights—that show the highest keyword coverage in both datasets. We believe this may be because authors tend to use more general rather than specialized terms when selecting keywords to describe the overall content of their papers. For instance, in a computer science paper proposing a new algorithm, the specific name of the algorithm might not appear in the keywords. Instead, the keywords are more likely to include the underlying principles or the tasks addressed by the algorithm, which are often presented in the general research introduction.

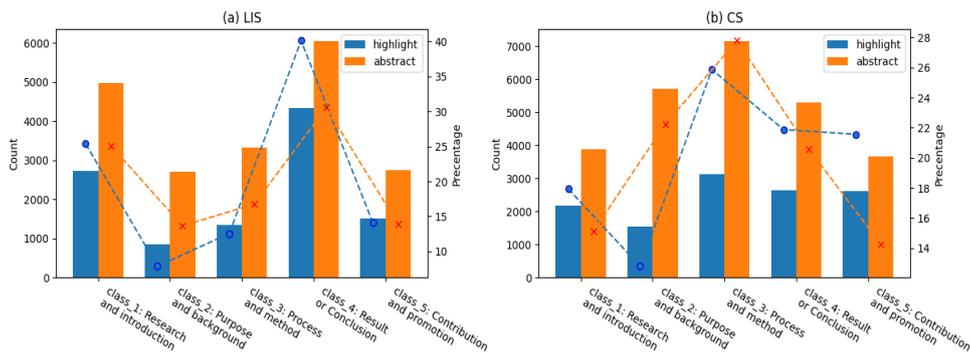

**Fig 7. Category distribution of highlights and abstract sentences**

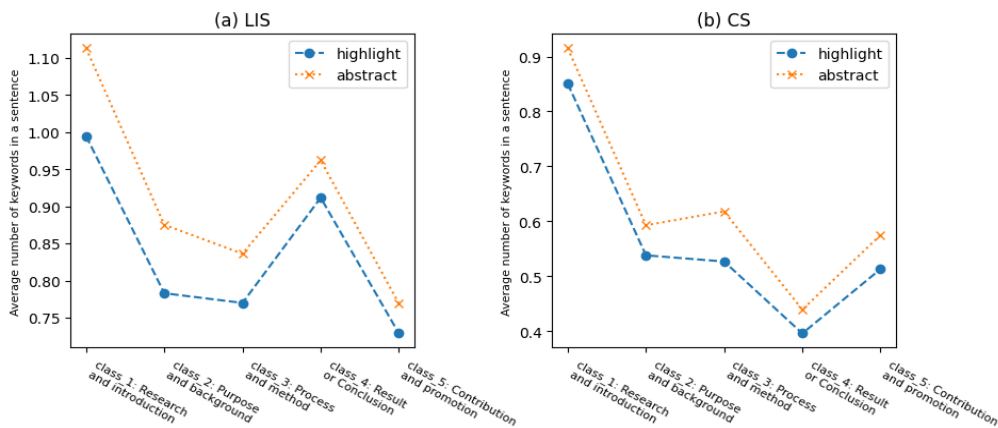

**Fig 8. Distribution of keywords in different types of sentences**

Considering the sentence type distribution in Figure 7 and the keyword coverage distribution in Figure 8, we conclude that using highlights for extracting keywords related to research results or conclusions may be more effective. Although these sentence types exhibit lower keyword coverage in highlights, their proportion is higher. This suggests that in disciplines focused on result analysis and the explanation of phenomena, integrating highlights with abstract can significantly enhance the performance of unsupervised extraction models. This inference is supported by the data in Tables 5 and 6, where the improvement in various metrics after incorporating highlights is more pronounced in the LIS dataset.

## 6. Discussion

### 6.1 Insights from Extraction Performance and Model Behaviors

In this study, we investigated whether incorporating highlight information enhances the performance of traditional, abstract-based unsupervised keyword extraction. Through comprehensive comparisons across various models and datasets, we found that while highlights generally yield performance gains, the magnitude and statistical significance of these improvements depend heavily on both the integration strategy and the choice of extraction model. Specifically, regarding integration strategies, we observed that semantically filtering irrelevant information from the abstract based on the highlights—and subsequently concatenating the filtered abstract with the highlights—outperforms using either the abstract or the highlight alone. However, this improvement is not statistically significant across most models and evaluation metrics. This lack of significance suggests that our filtering process may inadvertently introduce information loss by discarding keywords that, while not explicitly featured as "highlights" by the authors, still represent crucial thematic or background information. As supported by our analysis in Section 5.1, the abstract encompasses a broader breadth of keywords than the highlights. Therefore, an interesting avenue for future research is to develop more refined semantic analysis techniques to identify the characteristics of these abstract-exclusive keywords, allowing for a more precise aggregation of valuable information for the keyword extraction task.

Beyond semantic filtering, our comparison of direct concatenation strategies (Abstract+Highlight [A+H] and Highlight+Abstract [H+A]) revealed distinct performance variations, even though both strategies yielded statistically significant improvements in the vast majority of cases. As shown in Table 5 for the LIS dataset, the H+A strategy outperformed A+H across nearly all models and metrics. Conversely, in the CS dataset (Table 6), the advantage of the A+H strategy was primarily confined to the $F_1@5$ metric; when extracting a larger number of keywords (e.g., $F_1@10$ or $F_1@15$), H+A remained superior. Previous research has established that candidate word position is a critical feature in unsupervised keyword extraction, with words appearing earlier in the text having a higher probability of being actual keywords. Drawing on this, we hypothesize that because highlights possess a higher information density—with a more concentrated presence of keywords compared to the abstract—placing the highlight at the beginning of the input text (H+A) effectively leverages this positional bias to boost overall extraction performance. However, recognizing that the impact of positional information may vary according to the specific writing paradigms and conventions of different disciplines, future studies should extend this exploration to a wider range of academic fields to determine whether this positional advantage is a universal rule or a discipline-specific characteristic.

Aside from the choice of integration strategy, comparing the extraction models themselves provided further valuable insights. Among traditional extraction methods, approaches based on Pre-trained Language Models (PLMs) consistently outperformed legacy graph-based models, demonstrating substantial improvements across all metrics. This reaffirms the well-documented effectiveness of semantic understanding in enhancing text-based research tasks within the Information Retrieval and Natural Language Processing communities. A far more striking contrast, however, is observed between Large Language Models (LLMs) and traditional models. As detailed in the comparisons between Sections 4.2 and 4.4, the integration of LLMs resulted in a massive leap in extraction performance—a magnitude of improvement that dwarfs the incremental gains previously achieved through multi-source data integration or intricate feature engineering. This paradigm shift underscores the necessity for researchers to actively embrace and appropriately integrate LLMs into current methodologies. Simultaneously, it reveals that traditional approaches have severely underutilized the information embedded in the input text. Consequently, a compelling direction for future work is to conduct a granular, lexical-level comparative analysis between the keywords extracted by LLMs and those extracted by traditional models. By examining exactly which types and segments of keywords are uniquely captured by LLMs from the abstract and highlight texts, we can begin to demystify the "black-box" nature of these models. Ultimately, such semantic-level analyses will deepen our fundamental understanding of both the keyword extraction task itself and the inherent informational architecture of academic texts.

However, despite the massive baseline superiority of LLMs discussed above, a notable exception arises when applying the aforementioned information integration strategies to them. As detailed in Section 4.4, the application of various fusion strategies to the GPT-4o model yielded statistically insignificant improvements across most metrics on the CS dataset; in certain configurations, the fused inputs even underperformed compared to the abstract-only baseline. Similarly, the performance gains observed on the LIS dataset were largely insignificant in most cases for this model. Given the rapid evolution of Large Language Models and the fact that our study only evaluated two specific LLMs on this task, it is challenging to pinpoint the exact cause of this discrepancy. It remains unclear whether this anomaly is driven by specific model architectures (for instance, LLaMA-3 still demonstrated significant improvements when fusing abstracts and highlights), specific model versions (i.e., whether future iterations of GPT models would still exhibit this issue), or the specific prompt design we employed. The latter, in particular, has been widely recognized as a critical factor influencing performance in many studies applying LLMs to scientific tasks. We intend to explore these issues more thoroughly in future work. Beyond comparing a broader range of LLMs, a key direction will be investigating how to tailor prompt designs to the distinct characteristics of highlights and abstracts (drawing upon our analyses in Sections 5.1 and 5.2). Such tailored prompt engineering could be essential to maximizing the capability of LLMs in extracting keywords effectively when fusing these two distinct types of information.

## 6.2 The Increasing Prevalence of the Highlight Structure

Although the highlight field is primarily found in journal articles published by Elsevier, our investigation shows that an increasing number of publishers and journals have begun to require authors to provide highlights or highlight-like summaries at the time of submission. Table 16 presents several examples across publishers and subject areas, including psychology, biomedical sciences, chemistry, and engineering. In some journals, this section appears under different names—for example, Summary in Developmental Science and Learned Publishing, or Key Points in the American Journal of Clinical Pathology. However, our review of submission guidelines and published articles reveals that these structures serve essentially the same function: concisely listing the main findings and contributions of the paper in bullet-point form. In addition, Taylor & Francis has launched a pilot initiative called Key Policy Highlights (KPE)[9], which encourages authors to summarize the key policy implications of their research in order to facilitate communication between the scientific community and broader non-scientific audiences. A total of 36 journals under Taylor & Francis are participating in this pilot project.

While many publishers have not adopted standardized requirements as Elsevier has for all their journals, the above evidence indicates that the highlight structure is becoming increasingly common and more widely recognized and accepted among scholars. This trend reflects broader changes in the scientific publishing landscape. The surge in the number of academic papers, along with the rise of open access and diverse dissemination channels, has allowed wider audiences to access research outputs but has also made it more difficult for individual studies to be discovered. As Pottier et al. (2024) argue, traditional writing conventions for titles, abstracts, and keywords may constrain indexing practices in digital databases, thereby limiting the discoverability of scholarly publications. Against this backdrop, the introduction of new publishing structures such as structured abstracts (Roux & Burke, 2024), graphical abstracts (Lee et al., 2023), and highlights not only enables readers to quickly grasp the key points of a study during search and retrieval—thus improving research efficiency—but also helps authors increase the visibility of their work and foster scientific exchange.

Therefore, although the present study is based primarily on Elsevier articles, we anticipate that the applicability of our findings will extend more broadly as highlights and similar structures continue to proliferate in the future.

Table 16. Journals Requiring Author-Provided Highlights Across Publishers

| Publisher | Journal | Structure Name |
|---|---|---|
| Wiley | Developmental Science | Summary |
| | Clinical & Experimental Allergy | Summary |
| | Learned Publishing | Summary |
| | Health Information and Libraries Journal | Key messages |
| | Bioelectromagnetics | Summary |
| | Alzheimer's & Dementia | Highlights |
| Springer | Journal of Child and Family Studies | Highlights |
| | Environmental Process | Highlights |
| | SN Applied Sciences | Articles Highlights |
| | Biochar | Highlights |
| Oxford | International Journal of Epidemiology | Key Messages |
| | Briefings in Functional Genomics | Key points |
| | American Journal of Clinical Pathology | Key points |
| Cambridge University Press | Research Synthesis Methods | Highlights |

## 6.3 Scientometric Significance of the Study

Exploring the content of academic papers is a central concern in scientometrics, as textual elements serve as the foundation for mapping knowledge domains, identifying research fronts, and tracing the dynamics of scientific development. Previous studies have examined various aspects such as paper abstract (Sharma & Harrison, 2006), document structure (Sollaci & Pereira, 2004), future work sentences (Zhu et al., 2019), innovation evaluation elements (Yin et al., 2023), and even reference texts (Shibayama et al., 2021). However, research on highlights remains relatively limited.

---

[9] Taylor & Francis Author Services – *Key Policy Highlights*. Available at: https://authorservices.taylorandfrancis.com/publishing-your-research/writing-your-paper/key-policy-highlights/

As a concise, bullet-point representation of a paper's main contributions and innovations, highlights play an important role in improving reading efficiency and supporting the rapid dissemination of scientific knowledge. Evaluating the potential impact and significance of scientific publications has long been a key focus in scientometric research. While citation-based indicators and journal-level metrics have traditionally been the foundational tools for such evaluations, scholars are increasingly utilizing deep textual signals to judge the substantive contribution and novelty of a study. Highlights, by distilling the core ideas and contributions into an accessible format, enable researchers to quickly grasp the essence of a paper and assess its relevance to their own information needs. This not only facilitates more efficient scientific communication but also reduces the likelihood that valuable research is overlooked due to lengthy abstracts or full-text content.

Beyond the intrinsic importance of highlights in scholarly communication, this study demonstrates their value in a traditional scientometric task—keyword extraction. By integrating highlights with abstract information, we show that highlights can enhance the performance of unsupervised keyword extraction models. Because highlights summarize the essential content of a paper, their incorporation also creates opportunities for scientometric analyses such as innovation knowledge discovery, contribution-focused research evaluation, and research topic evolution modeling. As highlight-like structures continue to gain wider adoption across publishers and disciplines, their integration into scientometric workflows may further support more accurate topic modeling, co-word analysis, and the assessment of scientific contributions.

Overall, our findings enhance researchers' awareness of the importance of highlight texts and provide both empirical data and theoretical foundations for future studies involving highlights. More broadly, the results underscore that scientometric research can benefit substantially from deeper exploration of the structural and textual components of scholarly articles.

## 6.4 Implication

### 6.4.1 Theoretical implications

In the process of integrating abstract and highlights information, we propose a method for filtering abstract sentences based on their semantic similarity to highlights. We analyze how retaining different numbers of sentences from the filtered abstract affects model performance. Our findings indicate that even when only a small portion of the original abstract is retained, similar extraction performance can be achieved compared to using the full abstract. This suggests that abstract contain a significant amount of irrelevant information for keyword extraction and offers valuable insights and methods for researchers aiming to better understand and utilize abstract information for keyword extraction.

Ingwersen (1996) proposed the polyrepresentation theory to address the challenge that a single representation often fails to sufficiently capture the relevance between information objects and users' retrieval intentions. The theory argues that different representations reveal complementary aspects of relevance, and when these representations overlap, the likelihood of true relevance increases. Consequently, aggregating information from multiple representations can improve the effectiveness of tasks such as information retrieval and information extraction. This theoretical perspective resonates with recent advances in multimodal representation learning, where integrating complementary information from various modalities (e.g., text, speech, images, video) enables models to access richer semantic signals and achieve stronger performance in downstream tasks such as keyword extraction and entity recognition. In this study, although we focus solely on textual data, we exploit two distinct representations of research articles: highlights and abstracts. These two forms partially overlap in keyword coverage and textual content, yet differ in emphasis and expression. Such complementarity provides more diverse semantic cues for downstream tasks. When fused, highlights and abstracts enable the model to capture the core content of papers more comprehensively. The analysis in Section 5.2 further shows that the two text types differ significantly in sentence-type distributions, underscoring their distinct informational emphases. This provides empirical evidence for applying polyrepresentation theory within a single modality and theoretical support for why highlights can enhance the performance of unsupervised keyword extraction based on abstracts.

### 6.4.2 Practical implications

Recent work has introduced an unsupervised paradigm for keyword extraction that first selects salient sentences and then derives keywords from them, showing clear advantages on long documents (Zhang et al., 2023b). Our findings reinforce this paradigm: highlights, which summarize a paper's core contributions, contain concentrated keyword information and substantially improve extraction performance compared with abstracts alone. This suggests that effective keyword extraction may rely less on processing all textual content and more on leveraging sentence-level salience. Therefore, although our experiments are limited to papers with highlights, the results provide broader implications for keyword extraction in general corpora.

In this study, we compare several methods for integrating abstract and highlights information, finding that simply concatenating the two texts as input to the model yields better extraction performance than using either text alone. Zhang et al. (2022a) similarly employed a straightforward text concatenation approach in their study on improving keyword extraction by incorporating title information from references. However, unlike Zhang et al., we further explore the impact of concatenation order on extraction results, as previous research suggests that words appearing earlier in a text are more likely to be identified as keywords. Although adjusting the concatenation order did not significantly affect extraction performance in our experiments, we believe that future researchers should still consider this factor when conducting similar studies on keyword extraction involving the integration of multiple information sources.

Aiming to explain why integrating these texts as input leads to better performance in unsupervised keyword extraction tasks compared to using a single information source, this study examines the differences between abstract and highlights in terms of keyword content and sentence type distribution. We find that, compared to abstract, highlights have lower keyword coverage due to their shorter length. However, some highlights contain keywords that do not appear in the abstract, providing the keyword extraction model with additional candidate terms and potentially raising the performance ceiling. Additionally, there are significant differences in the distribution of sentence types between abstract and highlights. Highlights tend to include more descriptions of the study's conclusions and contributions. This suggests that for papers that are result-oriented and focus on analyzing research conclusions and phenomena, using highlights for keyword extraction might be more effective.

The analysis above deepens our understanding of the complementary roles of abstract and highlights. In future keyword extraction research, it may be valuable to explore how to design and select more appropriate information integration mechanisms or extraction models that leverage the strengths of these two types of texts.

### 6.5 Limitation

This study has several limitations: (1) Highlights are not present in all journal papers. Our experiments are conducted using only those academic papers that include this structure. Although the results indicate that integrating highlights with abstract can improve keyword extraction performance, the applicability of the method is limited. (2) The exploration of methods for integrating abstract and highlights is somewhat superficial. We only consider simple text concatenation and content filtering based on semantic similarity. We do not explore more suitable integration methods based on the specific characteristics of the content, nor do we investigate different integration approaches tailored to the requirements of specific extraction models. (3) While we conduct a preliminary investigation into the reasons for changes in model performance before and after integrating highlights information, the underlying mechanisms are not fully explained. Unsupervised extraction involves two stages: candidate word extraction and candidate word importance ranking. This study focuses only on the candidate word extraction stage, which is relatively straightforward to analyze. We do not examine how integrating abstract, highlights, or both as input affects the criteria for ranking candidate words, which directly influences keyword extraction results.

## 7 Conclusion and Future Works

This study investigates the use of academic paper highlights in unsupervised keyword extraction, focusing on how integrating information from both abstract and highlights can enhance extraction performance compared to using a single source. We explore various methods of information integration and conducted experiments using papers from the fields of LIS and CS. The results indicate that concatenating abstract and highlights as input leads to better model performance than using abstract alone. To further understand the reasons behind these performance changes, we compare keyword coverage between abstract and highlights, discovering that highlights often provide keywords not found in abstract. Additionally, sentence classification reveals that highlights contain a higher proportion of sentences describing research conclusions or contributions, making them more suitable for extracting keywords related to research results. In summary, our study not only enriches research on keyword extraction itself but also, by investigating the significance of highlights as a structural element in academic papers, encourages greater attention from researchers and promotes deeper exploration of their connections with other scientometric tasks, including innovation knowledge discovery and the identification of key research content.

Future work will focus on the following areas: (1) Advancing the Utilization of LLMs in Keyword Extraction; We plan to compare the keywords extracted by LLMs and traditional models to understand the mechanisms behind their massive performance leap. Furthermore, given their prompt sensitivity, we will design tailored prompts leveraging the distinct characteristics of highlights and abstracts to maximize extraction performance. (2) Exploring Automatic Generation of Highlights for Papers Lacking This Feature; To extend the applicability of our methods and conclusions, we will investigate how to automatically generate highlights for papers that do not have them. We aim to create the necessary corpus for subsequent keyword extraction research based on highlights. (3) Investigating the Focus of Highlights Across Different Disciplines and Subfields; This study find that the distribution of sentence types in highlights varies between fields, which can impact keyword extraction performance. By exploring the focus of highlights descriptions across different disciplines, we hope to better guide keyword extraction research based on abstract and highlights within specific fields.

# Acknowledgements

This work was supported by the National Social Science Fund of China (No. 24CTQ027).

# Ethics declarations

### Conflict of interest
The authors have no competing interests to declare that are relevant to the content of this article.